\documentclass[review,3p]{elsarticle}

\biboptions{square,comma,sort&compress}

\usepackage{graphicx}
\usepackage{epstopdf}
\DeclareGraphicsExtensions{.eps}

\usepackage[figoff,printfigures,figmark]{figcaps}

\usepackage{mymathstyle}

\usepackage{url}

\usepackage{lineno}

\makeatletter
\def\ps@pprintTitle{%
   \let\@oddhead\@empty
   \let\@evenhead\@empty
   \def\@oddfoot{\reset@font\hfil\thepage\hfil}
   \let\@evenfoot\@oddfoot
}
\makeatother

\begin{document}

\begin{frontmatter}
\title{Multitaper Spectral Analysis of Neuronal Spiking Activity Driven by Latent Stationary Processes}
\author[ece]{Proloy~Das}
\ead{proloy@umd.edu}
\author[ece,isr]{Behtash~Babadi\corref{cor1}}
\ead{behtash@umd.edu}
\cortext[cor1]{Corresponding author}
\address[ece]{Department of Electrical and Computer Engineering}
\address[isr]{Institute for Systems Research\\ University of Maryland, College Park, MD 20742, USA}
\begin{abstract}
Investigating the spectral properties of the neural covariates that underlie spiking activity is an important problem in systems neuroscience, as it allows to study the role of brain rhythms in cognitive functions. While the spectral estimation of continuous time-series is a well-established domain, computing the spectral representation of these neural covariates from spiking data sets forth various challenges due to the intrinsic non-linearities involved. In this paper, we address this problem by proposing a variant of the multitaper method specifically tailored for point process data. To this end, we construct auxiliary spiking statistics from which the eigen-spectra of the underlying latent process can be directly inferred using maximum likelihood estimation, and thereby the multitaper estimate can be efficiently computed.  Comparison of our proposed technique to existing methods using simulated data reveals significant gains in terms of the bias-variance trade-off.
\end{abstract}


\end{frontmatter}
\section{Introduction}
Spectral analysis techniques are among the most important tools for extracting information from time-series data recorded from naturally occurring processes. Examples include speech \citep{Quatieri:08}, image and video \citep{lim1990two}, and electroencephalography (EEG) \citep{Buzsaki:09} data. Due to the exploratory nature of most of these applications, nonparametric techniques based on Fourier methods and Wavelets are among the most widely used. In particular, the multitaper method excels among the available nonparametric techniques due to both its simplicity and {control over the bias-variance trade-off by means of bandwidth adjustment}\citep{mtm, walden2000unified, reviewmtm}. This technique has been successfully utilized in the analyze of EEG data \citep{PSseizures,freqseizures,reviewmtm}.

The advent of large-scale invasive recording technologies from the brains of animals and humans, such as multi-electrode arrays and electrocorticography (ECoG), has resulted in abundant pools of neuronal spiking data, which often exhibit oscillatory features \citep{ward2003synchronous}. Characterizing the properties of these oscillations with high spectral resolution is crucial to understanding their role in cognitive functions. 

Most existing spectral analysis techniques, however, are designed for continuous-time data and cannot be readily applied to binary spiking data (See, for example, \citep{DasDBMT} and \citep{Kim201702877}). There have been efforts aimed at addressing this challenge, which consider the periodogram of smoothed spike trains using kernel methods as the spectral representation \citep{chalk2010114,lewand2011,park2013kernel}. Another strand of results are based on the theory of point processes, which has been widely used in recent years to model and analyze the statistical properties of binary spike trains \citep{truccolo1074,panin2004,wu2017gaussian,xu2017sparse}. These techniques relate the Conditional Intensity Function (CIF) or the spiking rate of a point process governing the spiking statistics to intrinsic and external neural covariates using state-space models. Then, the spectrum of the estimated CIF is characterized using standard nonparametric or parametric techniques\citep{ssm,djuric2008target,brown2004}.

Despite their relative success in application, these methods have several shortcomings from a theoretical perspective. First, it is known that direct smoothing of the signal using kernels or indirect smoothing using state-space models, results in the distortion of the spectrum \citep{percival1993}. Second, time-domain smoothing alleviates the variance of the estimates at the cost of increasing the bias. On the other hand, existing techniques which avoid time-domain smoothing (e.g., \citep{miran2017robust}) may exhibit high variability. Third, these modeling frameworks often require a priori information (e.g., sparsity) or may suffer from model mismatch (e.g., overly-smoothed state estimates).

To address these issues, in this paper we introduce a novel multitaper spectral analysis method, which we call the Point Process Multitaper Method (PMTM), to be directly applied to binary data. To this end, we generate auxiliary spiking statistics which correspond to the tapered versions of the CIF, which are then used to independently estimate the eigen-spectra of the tapered CIFs via the Maximum Likelihood (ML) procedure. The multitaper spectral estimate is formed by averaging the corresponding eigen-spectral estimates. Our approach distinguishes itself from existing work by \emph{directly} estimating the spectra from binary observations via a novel adaptation of multitaper analysis, with no recourse to smoothing or need for a priori information. We demonstrate the performance of PMTM using simulated spike trains driven by an autoregressive (AR) process. Our results reveal substantial gains achieved by PMTM as compared to existing nonparametric techniques, in terms of the bias-variance trade-off. 

 
\vspace{-2mm}
\section{Problem Formulation}
\label{sec:ProbFrom}
Let $N(t)$ and $H_t$ denote the point process representing the number of spikes and spiking history in $[0,t)$, respectively, where $t  \in [0,T]$ and $T$ denotes the observation duration. The CIF of a point process $N(t)$ is defined as:
\begin{equation}
\lambda(t|H_t) := \lim_{\Delta \rightarrow 0} \frac{P[N(t+\Delta)-N(t) = 1|H_t]}{\Delta}.
\end{equation}  
 To discretize the continuous process, we consider time bins of length $\Delta$, small enough that the probability of having two or more spikes in an interval of length $\Delta$ is negligible. Thus, the discretized point process can be modeled by a Bernoulli process with success probability $\lambda_k : = \lambda(k\Delta|H_{k}) \Delta$, for $1 \le k \le K$, where $K := T/\Delta$ and is assumed to be an integer with no loss of generality. Note that $\lambda_k$ forms the CIF of the discretized process, which we refer to as CIF hereafter for brevity. Let $n_k \in \{0,1\}$ be the number of spikes in bin $k$, for $0 \le k \le K$. Our objective is to estimate the Power Spectral Density (PSD) of the CIF from the observed spike train $\{n_k\}_{k = 1}^{K}$, under the assumption that the CIF is a second-order stationary process.

More generally, we consider an ensemble of $L$ neurons or $L$ trials from a single neuron driven by the same CIF, and denote the observed spike trains by {$\mathcal{D}:=\big\{n_k^{(l)}\big\}_{k = 1,l = 1}^{K,L}$}. When considering an ensemble of $L$ neurons, this setting may only be valid for neuronal recordings from a small area of cortex using multi-electrode arrays (See, for example \citep{lewis2012rapid}), and when considering $L$ trials from the same neuron, it is assumed that all trials pertain to the same stimulus (See, for example \citep{sheikhattar2016recursive}). Nevertheless, in many other cases of interest, one has only access to a single trial from a single neuron, which makes spectral estimation an even more challenging problem. We model the CIF using a second order zero-mean stationary random process, $\{x_k\}_{k = 1}^{K}$, which by virtue of Spectral Representation Theorem \citep{percival1993} admits a Cram\'{e}r representation \citep{Loeve63} of the form:
\begin{equation}
\label{eq:cramer}
x_k = \int_{-\frac{1}{2}}^{\frac{1}{2}} e^{i2\pi f k} dz(f),
\end{equation}
where $dz(f)$ is a {complex-valued} orthogonal increment process and the PSD, $S(f)$ of the process is defined as: $S(f)df = \mathbbm{E}[|dz(f)|^2]$. Finally, we use a linear link for the CIF so that model can be summarized as
\begin{align}\label{eq:model}
 \lambda_k  =  \mu + x_k, \ \ \ n_k^{(l)} \sim \textrm{\small \sf Bernoulli}(\lambda_k),
\end{align}
for $\ 1 \le k \le K$ and  $1 \le l \le L$. The choice of the linear link, as opposed to more natural links such as the logistic function, is for the sake of simplicity of the auxiliary data generation process that will be described in Section \ref{sec:aux}. Acknowledging the non-linearity of the model (\ref{eq:model}) and availability of only a \textit{finite} number of samples, we consider a piece-wise continuous approximation to the PSD, i.e, $dz(f)$ is constant over the intervals $[\frac{m-1}{2N},\frac{m}{2N})$, for large enough $N$, for $m=1,2,\cdots,N$. This enables us to express $dz(f) = (a_m + ib_m)df$ for $f \in [\frac{m-1}{2N},\frac{m}{2N})$, where $a_m$ and $b_m$ are random variables for $m=1,2,\cdots,N$ \citep{miran2017robust}. Invoking the conjugate symmetry of $dz(f)$ for real valued $\{x_k\}_{k = 1}^{K}$, Eq. (\ref{eq:cramer}) can be written as
\begin{align}
\label{eq:approximation}
x_k\!=\!\sum_{m = 1}^{N} \frac{2}{N}\Big[a_m \cos \frac{\pi(m\!-\!1)}{N} - b_m \sin \frac{\pi(m\!-\!1)}{N}\Big],
\end{align}
with a PSD of $S(f) = \frac{1}{N}\mathbbm{E}[a_m^2 + b_m^2]$ for $f \in [\frac{m-1}{2N},\frac{m}{2N})$.

Denoting $ \mathbf{x} := [x_1 , x_2 , \cdots , x_K]^\top $ and $\mathbf{z} := [a_1, a_2, b_2, \cdots , a_N , b_N]^\top$ and defining $\mathbf{A}$ as
\begin{align*}
{
\!\mathbf{A}\!\!:=\!\! \frac{2}{N}\!\!
\begin{bmatrix}
1       & \cos\!\frac{\pi}{N}   & -\sin\!\frac{\pi}{N}   & \dots  & \cos\!\frac{(N-1)\pi}{N}   & -\sin\!\frac{(N-1)\pi}{N} \\
1       & \cos\!\frac{2\pi}{N}  & -\sin\!\frac{2\pi}{N}  & \dots  & \cos\!\frac{2(N-1)\pi}{N}  & -\sin\!\frac{2(N-1)\pi}{N} \\
\vdots  &  \vdots				&			\vdots		& \ddots &  		\vdots			  &	\vdots					 \\
1       & \cos\!\frac{K\pi}{N}  & -\sin\!\frac{K\pi}{N}  & \dots  & \cos\!\frac{K(N-1)\pi}{N}  & -\sin\!\frac{K(N-1)\pi}{N}
\end{bmatrix}}\!,
\end{align*}
one can write (\ref{eq:approximation}) in the vector form as $\mathbf{x} = \mathbf{Az} $. By the orthogonality of the increment process $ dz(f)$, $z_i$'s are independent. We further assume that $z_i$, for $i=1,2,\cdots,2N-1$, follows a truncated zero-mean Gaussian distribution, with a density
\begin{align}
\frac{f_i(z_i) \mathbbm{1}[ -{\mu} \leq z_i \leq {\mu} ] }{\int  f_i(\xi) \mathbbm{1}[ -{\mu} \leq \xi \leq {\mu} ] d{\xi}} \ ,
\end{align}
where $f_i(\cdot)$ is the Gaussian density $\mathcal{N}(0, {\sigma_i}^2)$ and $\mathbbm{1}$ is the indicator function. This ensures that each entry of $\{x_k\}_{k = 1}^{K}$ is restricted to $[-\mu, \mu]$ for some positive scalar $\mu$ and by selecting $\mu \leq 1/2$, one can achieve $0 \leq \lambda_k \leq 1$, for all $1 \leq k \leq K $.
\section{The Point Process Multitaper Method}
The MTM is an extension of tapered PSD estimation, where the spectral estimate is computed by averaging several tapered PSD estimates corresponding to orthogonal tapers with optimal spectral leakage properties \cite{percival1993}. The set of tapers from the \textit{Discrete Prolate Spheroidal Sequences (dpss)} \citep{slepian78} provides excellent control over the bias-variance trade-off \citep{reviewmtm,mtm,walden2000unified}.
 
 Let $v_k^{j}$ is be the $k^\text{th}$ sample of the $j^\text{th}$ dpss sequence, for a given design bandwidth $W$, $k=1,2,\cdots,K$ and $j=1,2,\cdots,J$ {such that $J < \lfloor 2 K W\rfloor - 1$}. Given the time-series data $\big\{x_k\big\}_{k = 1}^K$, the 
 $j^\text{th}$ eigen-spectrum is given by: 
\begin{align}
\!\!\widehat{S}^{(j)}(f) :=& \bigg|\sum_{k=1}^{K}e^{-i2\pi f k}v_k^{(j)}x_{k}\bigg|^2 \text{  for } j = 1,2, \cdots J \label{eq:eigen-spectra}
\end{align}
from which the MTM PSD estimate can be computed as:
\begin{align}
\widehat{S}^{(\sf mtm)}(f) := \frac{1}{J}\sum_{j=1}^{J}\widehat{S}^{(j)}(f). \label{eq:mtm-spectra}
\end{align}
Due to the non-linear nature of the model (\ref{eq:model}), forming the MTM PSD estimate based on spiking data becomes non-trivial. In what follows, we indeed address this issue by devising a novel variant of MTM.
\begin{figure}[!htbp]
\centering
\includegraphics[width=0.6\columnwidth]{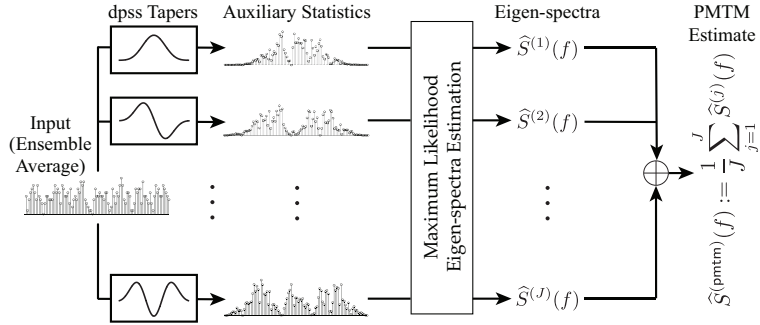}%
\caption{Schematic depiction of the proposed method. Stem plots show the ensemble average of the underlying spike trains.}
\label{fig_sim}
\end{figure}
\subsection{Generating Auxiliary Spiking Statistics}\label{sec:aux}
Given that $\{ \lambda_k\}_{k=1}^K$ is not directly observable, we instead generate auxiliary spiking statistics which would have been generated by the tapered CIFs, i.e., $\big\{v_k^{(j)}\lambda_{k}\big\}_{k = 1}^{K}$. For non-negative $v_k^{(j)}$ (e.g., for $j=1$), generating such modified spike trains is usually carried out using the thinning method \citep{cox1980}. The basic idea of the thinning method is to retain the original spikes with a probability determined by the ratio of the target and original CIFs. Noting that the ensemble average of the spike trains form a sequence of sufficient statistics for estimating the CIF, one can generate several realizations of thinned spike trains and consider their ensemble average as a sufficient statistic for the thinned process. It is not difficult to see that for non-negative tapers, the ensemble average converges in probability to the original spiking activity multiplied element-wise by the taper \citep{billingsley2008probability}. Thus, for the sake of robustness we use this limit as the ensemble average of the auxiliary spike trains.

Given that the dpss tapers take negative values (for $j > 1$), the aforementioned thinning method for constructing spike trains from a tapered CIF cannot be applied, as its na\"{i}ve application results in {negative-valued spikes}. To resolve this issue, we use the fact that the spike train is generated according to a Bernoulli process and therefore its complement given by $\check{n}^{(l)}_k := 1 - n^{(l)}_k$ has a CIF given by $ \check{\lambda}_k := 1 - \lambda_k = 1 - \mu - x_k $. Therefore, the \emph{non-negative} sequence
\begin{equation}
\label{eq:auxspike}
n^{(l,j)}_k = n_k^{(l)} v^{(j)}_k\mathbbm{1}[v^{(j)}_k \ge 0]\!-\!\check{n}^{(l)}_kv^{(j)}_k\mathbbm{1}[v^{(j)}_k < 0] 
\end{equation}
\noindent has a limit ensemble average corresponding to the CIF:
\begin{align}
\label{eq:auxCIF}
\lambda^{(j)}_k = \mu_k^{(j)} + x_kv^{(j)}_k,
\end{align} 
\noindent where $\mu_k^{(j)} = \mu v^{(j)}_k \mathbbm{1}[v^{(j)}_k\!\ge\!0]\!-\!(1\!-\!\mu)v^{(j)}_k \mathbbm{1}[v^{(j)}_k\!<\!0]$, for $j=1,2,\cdots,J$, and can be utilized as a sequence of sufficient statistics to estimate the spectral representation of the tapered process. Note that the non-negativity of the sequence $n_k^{(l,j)}$ follows from the definition of $\check{n}_k^{(l)}$. In addition, each taper is scaled by its maximum absolute value to ensure $n_k^{(l,j)} \le 1$, and the estimated eigen-spectra are accordingly rescaled. Using this approach, we can  generate the ensemble average of auxiliary spike trains for any taper regardless of its sign. It is noteworthy that while in principle this procedure can be extended to more general link functions, the generation of the corresponding auxiliary statistics may be more intricate. Fig. \ref{fig_sim} provides a visual summary of our proposed framework. The time-bandwidth product and the number of tapers are chosen following guidelines from the MTM literature \citep{mtm,reviewmtm}. 
\subsection{Maximum Likelihood Estimation of the Eigen-spectra} 
\label{emes}
Once the auxiliary spiking statistics $\mathcal{D}^{(j)} = \big\{n^{(l,j)}_k\big\}_{k = 1,l=1}^{K,L}$, $j = 1,2,\cdots,J$ are available, the eigen-spectra need to be estimated to construct the PSD. Given the modeling framework of Section \ref{sec:ProbFrom}, estimation of the $j^\text{th}$ eigen-spectrum reduces to estimating the parameters $\boldsymbol{\theta}^{(j)} := [{\sigma_1^{(j)2}}, {\sigma_2^{(j)2}}, \cdots, {\sigma^{(j)2}_{2N-1}}]^\top$, where ${\sigma_m^{(j)2}}$ is the variance of the random variable $z_i^{(j)}$ corresponding to the $j^\text{th}$ eigen-spectra, $i=1,2,\cdots, 2N-1$.  The ML estimate of the parameter $\boldsymbol{\theta}^{(j)}$ is given by:
\begin{align}
\label{eq:ML}
\widehat{\boldsymbol{\theta}}^{(j)}_{\sf ML} = \argmax_{\boldsymbol{\theta}^{(j)}} P(\mathcal{D}^{(j)}|\boldsymbol{\theta}^{(j)})
\end{align}
Note that expressing $P(\mathcal{D}^{(j)}|\boldsymbol{\theta}^{(j)})$ solely in terms of $\mathcal{D}^{(j)}$, i.e., eliminating $\mathbf{z}^{(j)} := [z_1^{(j)}, z_2^{(j)}, \cdots, z_{2N-1}^{(j)}]^\top$, introduces computational intricacies, which we avoid by using the Expectation-Maximization (EM) algorithm \citep{em1977} as our solution method. In what follows we drop the superscript $j$ for the sake of clarity, as the same procedure will be used for estimating each eigen-spectrum. If $\mathbf{z}$ is known, the \textit{complete} data log-likelihood of the observations $\mathcal{D}$ can be written as:
\begin{align}
\label{eq:com-ll}
 \log L(\boldsymbol{\theta}|\mathbf{z},\mathcal{D}) =\sum_{l=1}^{L}&\!\sum_{k=1}^{K}\!\!\bigg[n_k^{(l)} \log\frac{\mu_k\!+\!(\mathbf{Az})_k}{1\!-\!\big(\mu_k\!+\!(\mathbf{Az})_k\big)}\!+\!\log{\big(1\!-\!\big(\mu_k + (\mathbf{Az})_k\big)\big)}\!\bigg] \nonumber \\
& -\sum_{m = 1}^{2N-1} \Big(\log\int_{\!-\!\mu}^{\mu}\!f_m(\mathbf{\xi})d\mathbf{\xi}+\frac{z_m^2}{2\sigma_m^2}+\frac{1}{2}\log\sigma_m^2\Big)+{C},
\end{align}  
\noindent which could be efficiently maximized to estimate $\boldsymbol{\theta}$ (terms independent of $\boldsymbol{\theta}$ are denoted by $C$).
 
 Before presenting the EM algorithm, we note that $\int_{-\mu}^{\mu}f_m({\xi})d{\xi} = 1 - 2{\Phi}(\gamma_m) \approx 1$ for moderate values of $\gamma_m := \frac{\mu}{\sigma_m}$ and small enough $\sigma_m^2$. Thus, we drop this term henceforth to avoid unnecessary complexity. One may choose to work with this term included, at the expense of additional computational costs. At the $i^\text{th}$ iteration, we have: 
\subsubsection{E-step} 
Given $\boldsymbol{\theta}^{[i]}$, the Q-function is given by
\begin{align}
\label{eq:Estep1}
\mathbf{Q}(\boldsymbol{\theta}|\boldsymbol{\theta}^{[i]}) =-\sum_{m = 1}^{2N-1} \bigg(\frac{1}{2}\log\sigma_m^2+\frac{1}{2\sigma_m^2}\mathbbm{E}[z_m^2|\mathcal{D},\boldsymbol{\theta}^{[i]}]\bigg)+{C'},
\end{align}
which requires $f_{\mathbf{z}|\mathcal{D},\boldsymbol{\theta}^{[i]}}$ or samples from it, and can thus be computationally demanding to compute (terms independent of $\boldsymbol{\theta}$ are denoted by $C'$). Instead, we use the unimodality of the density and approximate it by a multivariate Gaussian density $\mathcal{N}(\boldsymbol{\mu}_{\mathbf{z}^{[i]}},\boldsymbol{\Sigma}_{\mathbf{z}^{[i]}})$ \citep{ssm,truccolo1074}. By invoking the fact that the mode and mean of a multivariate Gaussian density coincide and the Hessian of its natural logarithm is equal to $-\big({\boldsymbol{\Sigma}_{\mathbf{z}^{[i]}}\big)}^{-1}$, we get:
\begin{align}
\label{eq:mode}
\boldsymbol{\mu}_{\mathbf{z}^{[i]}} = \argmax_{\mathbf{z} \in D} \ \ \sum_{l=1}^{L}\sum_{k=1}^{K} \bigg[n_k^{(l)} \log\frac{\mu_k+(\mathbf{Az})_k}{1-(\mu_k+(\mathbf{Az})_k)}+ \log{(1\!-\!(\mu_k + (\mathbf{Az})_k))}\bigg] -\sum_{m = 1}^{2N-1}\frac{z_m^2}{2\sigma_m^2} \ ,
\end{align} 
and $\boldsymbol{\Sigma}_{\mathbf{z}^{[i]}}$ is given by the Hessian of the log-likelihood in (\ref{eq:com-ll}) evaluated at $\boldsymbol{\mu}_{\mathbf{z}^{[i]}}$. The maximization problem (\ref{eq:mode}) is concave over $D = \{\mathbf{z} \in \mathbb{R}^{2N-1}:{0} \leq \mu_k + (\mathbf{Az})_k \leq {1}, k = 1,2,\cdots, K\}$ and the Hessian is negative definite, so Newton-type method for bound-constrained optimization can be used to compute $\boldsymbol{\mu}_{\mathbf{z}^{[i]}}$ efficiently. We use a line-search method \citep{boyd2004convex}, which generates a sequence of iterates by setting $ \boldsymbol{\mu}^{[r+1]}_{\mathbf{z}^{[i]}} = \boldsymbol{\mu}^{[r]}_{\mathbf{z}^{[i]}} + \alpha^{[r]}\mathbf{d}^{[r]}$, where $\boldsymbol{\mu}^{[r+1]}_{\mathbf{z}^{[i]}}$ is a feasible approximation to the solution, $\alpha^{[r]}$ is the step-size and $\mathbf{d}^{[r]}$ is the Newton's step for that iteration. Then, $\boldsymbol{\Sigma}_{\mathbf{z}^{[i]}}$ can be computed by evaluating the Hessian of (\ref{eq:com-ll}) at $\mathbf{z} = \boldsymbol{\mu}_{\mathbf{z}^{[i]}}$, which allows $\mathbbm{E}[z_m^2|\mathcal{D},\boldsymbol{\theta}^{[i]}]$ to be calculated as $\big((\boldsymbol{\mu}_{\mathbf{z}^{[i]}})_m\big)^2 + \big(\boldsymbol{\Sigma}_{\mathbf{z}}^{[i]}\big)_{m,m}$.
\subsubsection{M-step}
The parameter vector $\boldsymbol{\theta}^{[i+1]}$ is updated by maximizing the expectation in (\ref{eq:Estep1}). Given that $Q(\boldsymbol{\theta}|\boldsymbol{\theta}^{[i]})$ is concave over the positive orthant, its unique maximizer is given by $\widehat{\boldsymbol{\theta}}_m^{[i+1]} = \mathbbm{E}[z_m^2|\mathcal{D},\boldsymbol{\theta}^{[i]}]$. 

Note that we have assumed $\mu_k^{(j)}$'s to be known. Since it is not the case for most practical purposes, we first estimate $\mu$ as $\hat{\mu}= \frac{1}{LK}\sum_{l,k = 1}^{L,K}n_k^{(l)}$ and compute $\mu_k^{(j)}$ in (\ref{eq:auxCIF}) using $\hat{\mu}$. We terminate the EM algorithm after a fixed large number of iterations or until some convergence criterion is met. A similar stopping rule for the maximization problem inside each EM step is used. We initialize $\boldsymbol{\theta}^{[0]}$ as an arbitrary vector in the positive orthant. Following the termination of the EM algorithm, the eigen-spectra are calculated as $\widehat{S}(0) = \widehat{\sigma}_{1}^2$ and $\widehat{S}(f_m) = \widehat{\sigma}_{2m}^2 + \widehat{\sigma}_{2m+1}^2$ for $f_m = \frac{m}{2N}$ and $m = 1,2, \cdots N-1$. Finally, the PMTM estimate can be computed using (\ref{eq:mtm-spectra}). Algorithm \ref{algo2} summarizes the proposed PMTM procedure.
\setlength{\textfloatsep}{10pt}
\begin{algorithm}[!b]
\label{algo2}
 \SetKwInOut{Input}{Input}
 \SetKwInOut{Output}{Output}
    \Input{Ensemble of neuronal spiking data, $\big\{n_k^{(l)}\big\}_{k = 1, l = 1}^{K,L} $ for $k = 1,2, ..,K$; {Design bandwidth}, $W$, such that $\alpha := KW \ge 1$; Number of tapers, $J$.}
    \Output{PMTM estimate, $\widehat{S}^{(\sf pmtm)}(f)$}
    Generate $J < \lfloor 2\alpha\rfloor$ dpss corresponding to data length $K$ and half time-bandwidth product $ \alpha $\; 
\For{$j = 1$ to $J$}{
  Generate $\big\{n^{(l,j)}_k\big\}_{k = 1}^K$, for $l = 1,2,\cdots,L$ \;
  Compute $\widehat{S}^{(j)}(f)$ using ML estimation\;
  }
 Compute $\widehat{S}^{(\sf pmtm)}(f) = \frac{1}{J}\sum_{j = 1}^{J}\widehat{S}^{(j)}(f)$
\caption{The Point Process Multitaper Method}
\end{algorithm}
\section{Simulation Results}\label{ExpRes}
\begin{figure}[!htbp]
\centering
\includegraphics[width=0.55\columnwidth]{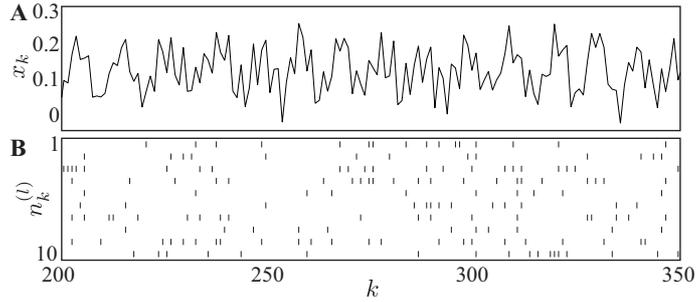}
\vspace*{-2mm}
\caption{(\textit{A}) A snapshot of the simulated AR process for $200 \le k \le 350$. (\textit{B}) Raster plot of the corresponding neuronal ensemble activity.}
\vspace*{-2mm}
\label{fig:SimData}
\end{figure}

We simulate $x_k$ as an AR$(4)$ process given by $\displaystyle x_k =  0.4152x_{k-1} - 0.0922x_{k-2} + 0.4170x_{k-3} - 0.8852x_{k-4} + 0.025\epsilon_k$, where $\epsilon_j$ is zero-mean i.i.d. Gaussian noise with unit variance. We compute the CIF as $\lambda_k = \mu + x_k$ for $\mu = 0.12$ {(truncated to $[0,1]$, if necessary)}, to generate the binary spiking activity for $K=512$ samples. A snapshot of one realization of this AR process and the raster plot of $L = 10$ spike trains are depicted in Fig. \ref{fig:SimData}\textit{A} and \textit{B}, respectively.
We apply PMTM to this simulated data and benchmark it against two existing methods: $(1)$ PSTH-PSD, where the PSD is computed by forming the MTM estimate of the ensemble peristimulus time histogram (PSTH), i.e., the average spike trains, and $(2)$ SS-PSD, where $x_k$ is first estimated using a state-space model $x_k = x_{k-1} + w_k$, followed by forming its MTM PSD estimate \citep{ssm}.

\begin{figure}[!htbp]
\centering
\includegraphics[width=0.55\columnwidth]{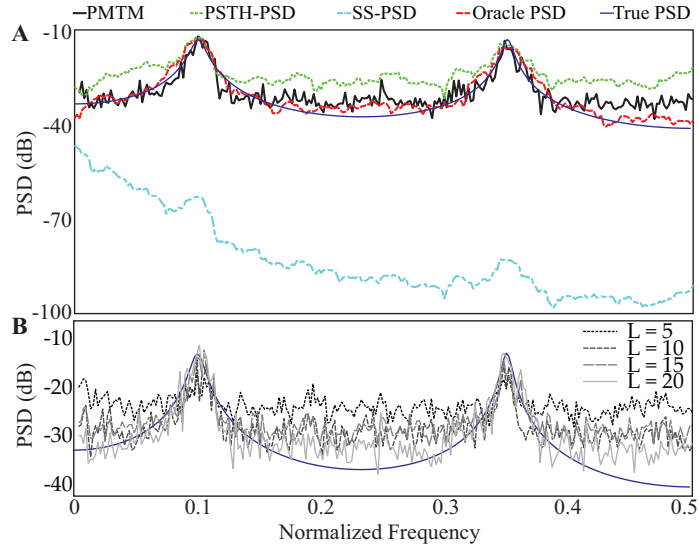}
\caption{{Comparison of the PSD estimates. (\textit{A}) PMTM, PSTH-PSD, SS-PSD, Oracle PSD (all using $\alpha = 5, J = 8$), and the true PSD. (\textit{B}) PMTM estimates for $L = 5,~10,~15$, and $20$.}}
\vspace*{-2mm}
\label{results}
\end{figure}
Fig. \ref{results}\textit{A} shows the PMTM (black), PSTH-PSD (green), SS-PSD (aqua) and the true PSD (blue) for the realization shown in \ref{fig:SimData} in log-scale. For comparison purposes, we have also included the MTM PSD estimate of $x_k$, assuming that an oracle has access to it (Oracle PSD, in red). We have used the first $8$ dpss tapers corresponding to $\alpha = 5$. As it can be observed from Fig. \ref{results}\textit{A}, PSTH-PSD suffers from high bias, though enjoying reduced variability, and the spectral peaks are difficult to distinguish from the background. On the other hand, SS-PSD suffers from model mismatch, as it over-smooths the CIF due to the usage of a state-space model, and as a result the spectral peaks are nearly absent in the estimate. The PMTM estimate, however, closely follows the true PSD by guarding against spectral leakage and producing a nearly unbiased estimate on par with the Oracle PSD estimate, though it exhibits some variability. In order to quantify these comparisons, we computed a normalized measure of MSE by averaging the squared-error of the PSD normalized by the true PSD values in the log-scale. The normalized MSE values ($\pm$ 2 STD) corresponding to $10$ different AR process realizations and $5$ different spike-train ensemble realizations are given in Table \ref{tab}, which corroborates our foregoing qualitative comparison. To ease reproducibility, we have deposited a MATLAB implementation of PMTM on the open source repository Github \citep{code}, which fully regenerates Fig. \ref{results}\textit{A}. 
\begin{table}[!th] 
\centering
\begin{tabular}[c]{c c c}
                     PMTM        &        SS-PSD  & PSTH-PSD       \\
\hline
   $0.4733 \pm 0.0072$  & $0.8164 \pm 7.9592\times 10^{-8}$ & $7.7772 \pm 2.0641$ \\
\end{tabular}
\caption{Normalized MSE Comparisons}
 \label{tab}
\end{table} 

Finally, Fig. \ref{results}\textit{B} examines the improvement of the PMTM estimates with respect to the ensemble size, for $L = 5,10,15$ and $20$. As $L$ increases, the PSD estimates improve, but with a seemingly saturating effect for $L \ge 10$. Note that while the average spiking rate of $\mu = 0.12$ is reasonable for some neurons (e.g., Parvalbumin-positive interneurons \citep{forli2018two}), often times neurons exhibit spiking rates as low as $\mu = 0.05$, for which the performance of all existing methods significantly degrades. Extension of our method for obtaining robust spectral estimates in the domain of low spiking rate and small number of trials is a future direction of research.
\section{Conclusion}
Spectral estimation of continuous time-series is a well-established domain, as hallmarked by the multitaper method known for its favorable control over the bias-variance trade-off. Computing the spectral representation of the neural covariates that underlie spiking activity, however, sets forth various challenges due to the intrinsic non-linearities involved. In this paper, we addressed this problem by proposing a multitaper method specifically tailored for binary spiking data, which we refer to as PMTM. We compared the performance of PMTM to that of two existing techniques using simulated data, which revealed significant gains in terms of estimation accuracy. The PMTM can be extended to a wide variety of binary data, such as rainfall and earthquake data, to extract spectral representations of the underlying latent processes.
%
%
\section*{Acknowledgment}
This work is supported in part by the National Science Foundation Awards No. 1552946 and 1807216.
%
\section*{References}
\bibliographystyle{elsarticle-num}
\bibliography{mybib}

\begin{thebibliography}{10}
\expandafter\ifx\csname url\endcsname\relax
  \def\url#1{\texttt{#1}}\fi
\expandafter\ifx\csname urlprefix\endcsname\relax\def\urlprefix{URL }\fi
\expandafter\ifx\csname href\endcsname\relax
  \def\href#1#2{#2} \def\path#1{#1}\fi

\bibitem{Quatieri:08}
T.~F. Quatieri, Discrete-time Speech Signal Processing: Principles and
  Practice, Prentice Hall, 2008.

\bibitem{lim1990two}
J.~S. Lim, Two-dimensional signal and image processing, Englewood Cliffs, NJ,
  Prentice Hall, 1990, 710 p.

\bibitem{Buzsaki:09}
G.~Buzsaki, Rhythms of the Brain, Oxford University Press, 2009.

\bibitem{mtm}
D.~J. Thomson, Spectrum estimation and harmonic analysis, Proceedings of the
  IEEE 70~(9) (1982) 1055--1096.
\newblock \href {http://dx.doi.org/10.1109/PROC.1982.12433}
  {\path{doi:10.1109/PROC.1982.12433}}.

\bibitem{walden2000unified}
A.~Walden, A unified view of multitaper multivariate spectral estimation,
  Biometrika 87~(4) (2000) 767--788.

\bibitem{reviewmtm}
B.~Babadi, E.~N. Brown, A review of multitaper spectral analysis, IEEE
  Transactions on Biomedical Engineering 61~(5) (2014) 1555--1564.
\newblock \href {http://dx.doi.org/10.1109/TBME.2014.2311996}
  {\path{doi:10.1109/TBME.2014.2311996}}.

\bibitem{PSseizures}
G.~Alarcon, C.~D. Binnie, R.~D. Elwes, C.~E. Polkey, Power spectrum and
  intracranial {EEG} patterns at seizure onset in partial epilepsy,
  Electroencephalography and clinical neurophysiology 94 5 (1995) 326--37.

\bibitem{freqseizures}
R.~Fisher, W.~Webber, R.~Lesser, S.~Arroyo, S.~Uematsu, High-frequency {EEG}
  activity at the start of seizures, Journal of Clinical Neurophysiology 9~(3)
  (1992) 441--448.

\bibitem{ward2003synchronous}
L.~M. Ward, Synchronous neural oscillations and cognitive processes, Trends in
  cognitive sciences 7~(12) (2003) 553--559.

\bibitem{DasDBMT}
P.~Das, B.~Babadi, Dynamic bayesian multitaper spectral analysis, IEEE
  Transactions on Signal Processing 66~(6) (2018) 1394--1409.
\newblock \href {http://dx.doi.org/10.1109/TSP.2017.2787146}
  {\path{doi:10.1109/TSP.2017.2787146}}.

\bibitem{Kim201702877}
S.-E. Kim, M.~K. Behr, D.~Ba, E.~N. Brown,
  \href{http://www.pnas.org/content/early/2017/12/15/1702877115}{State-space
  multitaper time-frequency analysis}, Proceedings of the National Academy of
  Sciences\href {http://dx.doi.org/10.1073/pnas.1702877115}
  {\path{doi:10.1073/pnas.1702877115}}.
\newline\urlprefix\url{http://www.pnas.org/content/early/2017/12/15/1702877115}

\bibitem{chalk2010114}
M.~Chalk, J.~L. Herrero, M.~A. Gieselmann, L.~S. Delicato, S.~Gotthardt,
  A.~Thiele, Attention reduces stimulus-driven gamma frequency oscillations and
  spike field coherence in {V1}, Neuron 66~(1) (2010) 114 -- 125.

\bibitem{lewand2011}
B.~C. Lewandowski, M.~Schmidt, Short bouts of vocalization induce long-lasting
  fast gamma oscillations in a sensorimotor nucleus, The Journal of
  Neuroscience 31~(39) (2011) 13936--48.

\bibitem{park2013kernel}
I.~M. Park, S.~Seth, A.~R. Paiva, L.~Li, J.~C. Principe, Kernel methods on
  spike train space for neuroscience: a tutorial, IEEE Signal Processing
  Magazine 30~(4) (2013) 149--160.

\bibitem{truccolo1074}
W.~Truccolo, U.~T. Eden, M.~R. Fellows, J.~P. Donoghue, E.~N. Brown, A point
  process framework for relating neural spiking activity to spiking history,
  neural ensemble, and extrinsic covariate effects, Journal of Neurophysiology
  93~(2) (2005) 1074--1089.
\newblock \href {http://dx.doi.org/10.1152/jn.00697.2004}
  {\path{doi:10.1152/jn.00697.2004}}.

\bibitem{panin2004}
L.~Paninski, Maximum likelihood estimation of cascade point-process neural
  encoding models, Computat. Neural Syst. 15 (2004) 243--262.

\bibitem{wu2017gaussian}
A.~Wu, N.~G. Roy, S.~Keeley, J.~W. Pillow, Gaussian process based nonlinear
  latent structure discovery in multivariate spike train data, in: Advances in
  Neural Information Processing Systems, 2017, pp. 3496--3505.

\bibitem{xu2017sparse}
S.~Xu, Y.~Li, T.~Huang, R.~H. Chan, A sparse multiwavelet-based generalized
  laguerre--volterra model for identifying time-varying neural dynamics from
  spiking activities, Entropy 19~(8) (2017) 425.

\bibitem{ssm}
A.~C. Smith, E.~N. Brown, Estimating a state-space model from point process
  observations, Neural Computation 15 (2003) 965--991.

\bibitem{djuric2008target}
P.~M. Djuric, M.~Vemula, M.~F. Bugallo, Target tracking by particle filtering
  in binary sensor networks, IEEE Transactions on Signal Processing 56~(6)
  (2008) 2229--2238.

\bibitem{brown2004}
E.~N. Brown, R.~E. Kass, P.~P. Mitra, Multiple neural spike train data
  analysis: state-of-the-art and future challenges, Nature neuroscience 7
  (2004) 456--461.

\bibitem{percival1993}
D.~B. Percival, A.~T. Walden, Spectral Analysis for Physical Applications,
  Cambridge University Press, 1993.

\bibitem{miran2017robust}
S.~Miran, P.~L. Purdon, E.~N. Brown, B.~Babadi, Robust estimation of sparse
  narrowband spectra from neuronal spiking data, IEEE Transactions on
  Biomedical Engineering 64~(10) (2017) 2462--2474.

\bibitem{lewis2012rapid}
L.~D. Lewis, V.~S. Weiner, E.~A. Mukamel, J.~A. Donoghue, E.~N. Eskandar, J.~R.
  Madsen, W.~S. Anderson, L.~R. Hochberg, S.~S. Cash, E.~N. Brown, P.~L.
  Purdon, Rapid fragmentation of neuronal networks at the onset of
  propofol-induced unconsciousness, Proceedings of the National Academy of
  Sciences 109~(49) (2012) E3377--E3386.

\bibitem{sheikhattar2016recursive}
A.~Sheikhattar, J.~B. Fritz, S.~A. Shamma, B.~Babadi, Recursive sparse point
  process regression with application to spectrotemporal receptive field
  plasticity analysis, IEEE Transactions on Signal Processing 64~(8) (2016)
  2026--2039.

\bibitem{Loeve63}
M.~Loeve, Probability Theory, D. Van Nostrand Co., London, 1963.

\bibitem{slepian78}
D.~Slepian, Prolate spheroidal wave functions, {F}ourier analysis, and
  uncertainty-{V}: the discrete case, Bell Syst. Tech. J. 57~(5) (1978)
  1371--1430.
\newblock \href {http://dx.doi.org/10.1002/j.1538-7305.1978.tb02104.x}
  {\path{doi:10.1002/j.1538-7305.1978.tb02104.x}}.

\bibitem{cox1980}
D.~R. Cox, V.~Isham, Point Processes, Chapman and Hall, 1980.

\bibitem{billingsley2008probability}
P.~Billingsley, Probability and measure, John Wiley \& Sons, 2008.

\bibitem{em1977}
A.~P. Dempster, N.~M. Laird, D.~B. Rubin, Maximum likelihood from incomplete
  data via the em algorithm, Journal of the royal statistical society. Series B
  (methodological) (1977) 1--38.

\bibitem{boyd2004convex}
S.~Boyd, L.~Vandenberghe, Convex optimization, Cambridge university press,
  2004.

\bibitem{code}
The Point Process Multitaper Method, Available on GitHub Repository:
  \url{https://github.com/proloyd/PMTM}, 2018.

\bibitem{forli2018two}
A.~Forli, D.~Vecchia, N.~Binini, F.~Succol, S.~Bovetti, C.~Moretti, F.~Nespoli,
  M.~Mahn, C.~A. Baker, M.~M. Bolton, O.~Yizhar, T.~Fellin, Two-photon
  bidirectional control and imaging of neuronal excitability with high spatial
  resolution in vivo, Cell reports 22~(11) (2018) 3087--3098.

\end{thebibliography}

\end{document}